\documentclass{article}
\usepackage{arxiv}

\usepackage[utf8]{inputenc}
\usepackage{todonotes}
\usepackage{array}
\usepackage{mathtools}
\usepackage{subcaption}
\usepackage{enumitem}
\usepackage{hyperref}
\usepackage{cleveref}
\usepackage{booktabs}
\usepackage{tikz}
\usepackage{dingbat}

\usepackage{titlesec}

\titlespacing\section{0pt}{4pt plus 4pt minus 2pt}{0pt plus 2pt minus 2pt}
\titlespacing\subsection{0pt}{4pt plus 4pt minus 2pt}{0pt plus 2pt minus 2pt}
\titlespacing\subsubsection{0pt}{4pt plus 4pt minus 2pt}{0pt plus 2pt minus 2pt}

\newcommand*{\horzbar}{\rule[.5ex]{2.5ex}{0.5pt}}

\newcommand{\system}{\textsc{Trevor}}

\usepackage{pifont}
\newcommand{\xmark}{\ding{53}}

\AtBeginDocument{%
  \providecommand\BibTeX{{%
    \normalfont B\kern-0.5em{\scshape i\kern-0.25em b}\kern-0.8em\TeX}}}

\begin{document}

\title{{\system}: A Mitigation Technique for SyncBleed in ZIPA}

\title{SyncBleed: A Realistic Threat Model and Mitigation Strategy for Zero-Involvement Pairing and Authentication (ZIPA)}


\author{
{\rm Isaac Ahlgren}\\
Loyola University Chicago\\Computer Science Department
\and
{\rm Jack West}\\
University of Wisconsin\\Computer Science Department
\And
{\rm Kyuin Lee}\\
University of Houston\\Department of Information Science Technology
\and
{\rm George K Thiruvathukal}\\
Loyola University Chicago\\Computer Science Department
\And
{\rm Neil Klingensmith}\\
Loyola University Chicago\\Computer Science Department
} 



\maketitle


\begin{abstract}
Zero Involvement Pairing and Authentication (ZIPA) is a promising technique for autoprovisioning large networks of Internet-of-Things (IoT) devices.
Presently, these networks use password-based authentication, which is difficult to scale to more than a handful of devices.
To deal with this challenge, ZIPA enabled devices autonomously extract identical authentication or encryption keys from ambient environmental signals.
However, during the key negotiation process, existing ZIPA systems leak information on a public wireless channel which can allow adversaries to learn the key.
We demonstrate a passive attack called SyncBleed, which uses leaked information to reconstruct keys generated by ZIPA systems.
To mitigate SyncBleed, we present {\system}, an improved key generation technique that produces nearly identical bit sequences from environmental signals without leaking information.
We demonstrate that {\system} can generate keys from a variety of environmental signal types under 4 seconds, consistently achieving a 90-95\% bit agreement rate across devices within various environmental sources.


\end{abstract}



\section{Introduction}

Internet of Things (IoT) devices require secure wireless communication channels to exchange data and coordinate with one another.
Without secured channels, IoT devices are susceptible to attacks such as man-in-the-middle~\cite{survdevpair, caveateptor} which jeopardize user privacy and trustworthiness of IoT devices.
To establish a secured channel, IoT devices must pair and establish a common cryptographic key.

Traditionally, IoT devices in a network individually pair with a central entity (such as a gateway or hub) that is assumed to be trusted.
This is usually done through a person intervening to type in a password.
However, human-mediated pairing is prone to many faults, particularly for IoT devices.

Zero involvement pairing and authentication (ZIPA) aims to alleviate the pain of key management for IoT devices by authenticating them without use of human generated passwords.
Devices authenticating to a ZIPA network validate their legitimacy by proving that they are located in the same physical space (i.e., office, home) at the same time. 
The devices generate authentication keys from ambient environmental contexts such as electromagnetic radiation, audio, voltage, etc, which are chosen to be observable only within the confines of a protected physical space. 
External observers record different environmental context and cannot recreate the key.

\begin{figure}
    \centering
    \includegraphics[width=\hsize]{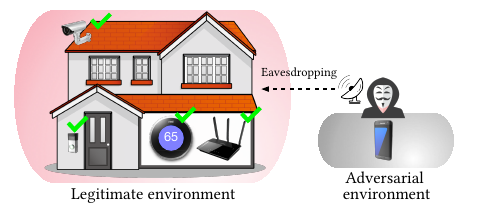}
    \caption{Threat model used for SyncBleed. Legitimate devices within the same environment authenticate, while the eavesdropping adversary tries to authenticate in different environment. SyncBleed provides an advantage to eavesdroppers by amplifying the quality of information they record from afar.}
    \label{fig:aliceandbob}
\end{figure}

Compared to traditional password-based authentication, ZIPA is more secure and easier to use. 
Because the devices autonomously authenticate themselves, users do not have to configure passwords on individual devices.
This improved usability promotes system's overall security because it allows the devices to autonomously and periodically rotate keys.


However, existing ZIPA protocols have a flaw that causes them to divulge information about keys they generate over an unsecured wireless channel.
In this work, we develop an attack called SyncBleed that targets the ZIPA key generation pipeline common to all ZIPA systems we are aware of.

ZIPA systems agree on a key by first sampling a time series of environmental signal, then synchronizing their sample buffers, and finally extracting a key from the synchronized buffers.
Without perfectly synchronized sample buffers, legitimate devices often fail to pair, as we demonstrate in \S\ref{sec:shiftintolerance}.
Indeed, the synchronization protocol used by most ZIPA systems is extremely reliable.
It works by sharing a snippet of the sampled environmental context over a public wireless channel.
The snippet allows the ZIPA devices to accurately align their buffers.
Although the synchronization snippet can be heard by external eavesdroppers, previous authors have assumed that this synchronization technique is safe because  the snippet is discarded and never directly used to compute the key.
This security model assumes that synchronization messages leak no information about the key or the legitimate space.


This assumption is valid in an ideal setting: legitimate devices on a ZIPA network are isolated from external adversaries by a barrier that does not allow environmental signals to pass.
A locked soundproof room, for example, could perfectly isolate audio from the outside world.
Malicious devices outside the isolated room cannot hear the environmental context inside, and they cannot influence the environmental context inside.
Without access to the environmental context, outside devices cannot generate a valid key.
The threat model of existing ZIPA systems presumes near-perfect isolation\cite{aerokey,proximate,cba-zipa,Schurmann-TMC13,voltkey}.

In this work we present an attack that allows us to learn the ZIPA key only by listening to synchronization messages on the unencryped public channel and recording environmental context from a distance (shown in \Cref{fig:aliceandbob}).
Our attack, which we call SyncBleed, targets the synchronization phase of ZIPA key generation.

SyncBleed gleans information from synchronization  messages to substantially narrow the key space.
In our testbed that uses the best existing ZIPA algorithms to generate keys in real time (\S\ref{sec:infoleak}), our SyncBleed attack could find about 50\% of the keys generated in less than a second. 


To mitigate SyncBleed, ZIPA protocols should avoid exchanging context-derived synchronization messages on a public wireless channel.
But removing synchronization messages presents a major challenge.
Existing network time synchronization protocols---most of which rely on probabilistic measurements of channel latency---do not reliably provide the microsecond-level accuracy needed for ZIPA systems~\cite{clocksync1,clocksyncreview}.
Precise time synchronization with peripherals like atomic clocks, GPS, or other hardware provide the near-perfect time synchronization necessary, but in practice they are too large or expensive to be deployed on low-cost IoT devices. 

To address this problem, we introduce {\system} (\textbf{T}ime shift \textbf{RE}sistant \textbf{VE}ctor Extract\textbf{OR}), a practical and general-purpose bit quantization technique that does not require precise time synchronization. 
{\system} does not exchange time synchronization messages and therefore does not suffer from the privacy, security, and speed limitations of other ZIPA protocols.
{\system} is fast: it can establish a shared key in under five seconds (\S \ref{sec:time_after_audio}).
Its bit quantization algorithm requires no communication between authenticating devices.
Because {\system} does not share any information during synchronization, its security is substantially improved over existing ZIPA protocols (\S \ref{sec:infoleak}).
{\system} is reliable: it achieves 90-95\% bit agreement rate between legitimate devices while rejecting all authentication attempts by the adversaries---competitive with the most recent ZIPA protocols~\cite{voltkey,aerokey,h2h,h2b}.
And most importantly, {\system} cannot be comprimised by SyncBleed.

Additionally, {\system} algorithm can be incorporated into existing ZIPA systems across different types of environmental signals.
In our evaluation, {\system} performs well on audio, electromagnetic, and voltage noise---three of the most commonly used types of environmental signal reported in related work on ZIPA~\cite{Ensemble,perils,Rowe-SenSys09,Schurmann-TMC13,soundproof,voltkey}.

{\system}'s basic approach is to authenticate using the frequency spectrum (an FFT) of environmental noise.
This FFT is invariant for small time shifts (within hundreds of milliseconds), provided there is overlap of measured signal between authenticating devices.
But within a signal's FFT, only a few of the components encode for common-mode information that is likely to be shared among multiple nearby devices.
A large portion of the signal is noise.
Choosing the frequency components that encode common-mode information is challenging.
Bands that encode useful common-mode signal are different for various types of environmental signals (audio, RF, voltage, vibration), and they change over time.

{\system} solves this problem by applying principal component analysis to the FFT of the environmental signal.
The principal components distill the strongest elements of the FFT, which are shared among nearby devices.
Shift invariance of the FFT followed by SNR amplification of principal component analysis is a good recipe for producing keys with high bit agreement rate that are not sensitive to small misalignments. 



\noindent
Our contributions include the following:
\vspace{-\topsep}
\begin{itemize}[leftmargin=*]
  \setlength{\parskip}{0pt}
  \setlength{\itemsep}{0pt plus 1pt}

    \item \emph{Threat Model:} We introduce a realistic threat model for ZIPA systems that assumes attackers can remotely measure environmental context in the legitimate space.

    \item \emph{SyncBleed:} We introduce a passive attack on ZIPA systems that exploits synchronization messages to guess the key.

    \item \emph{Mitigation:} We present {\system}, a practical, general-purpose mitigation for SyncBleed. {\system} is a bit quantization technique for ZIPA systems that tolerates misalignment. {\system} is not susceptible to SyncBleed because it does not exchange synchronization messages.
    
    \item \emph{System:} We deploy {\system} in a testbed and generate keys from environmental signals. {\system} tolerates misalignment time of up to a few hundred milliseconds and minimizes message exchange between devices.

    \item \emph{Evaluation:} We present a comprehensive evaluation {\system} on a diverse group of existing datasets, including audio, electromagnetic, and voltage.

\end{itemize}
\vspace{-\topsep}

\section{Background}
In this section our aim is to define and present the overall ZIPA pipeline, discuss similarities within ambient signals, and to describe our threat model.

\subsection{The ZIPA Pipeline}

\begin{figure}
    \centering
    \includegraphics[width=\hsize]{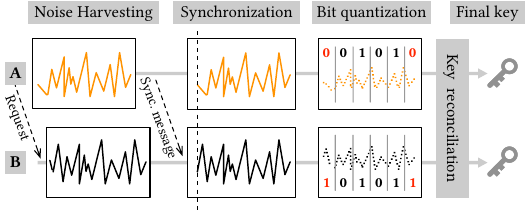}
    \caption{General pipeline of ZIPA between two authenticating
devices: $A$  and $B$ .}
    \label{fig:background}
\end{figure}

Most ZIPA systems assume that devices located within the same environment (i.e., home, office) are legitimate and any devices outside of such environment is considered as an adversary, as depicted in \Cref{fig:aliceandbob}.
ZIPA authentication between two devices ($A$ and $B$) generally involve four stages: noise harvesting, synchronization, bit quantization, and key reconciliation as illustrated in \Cref{fig:background}.

\begin{enumerate}[leftmargin=0cm,itemindent=.5cm,labelwidth=\itemindent,labelsep=0cm,align=left, noitemsep]

    \item \textbf{Noise Harvesting:} After receiving authentication request message from device $A$, two devices independently measure a sequence of samples from an environmental signal source.
    This is usually done by a microcontroller with an on-board analog-to-digital converter or other sensors.
    
    \item \textbf{Synchronization:} 
    Two devices time-align their signal measurements with device $A$ sending a short snippet of the measurement results (synchronization message) to device $B$ over a public unsecured channel.
    After receiving, device $B$ shifts the synchronization message over its own measured signal to obtain an identical starting point by searching for maximized Pearson correlation coefficient.
    
    \item \textbf{Bit quantization:} 
    Both devices independently convert the synchronized signal into a bit sequences to be later used as a key.
    Even though the devices $A$ and $B$ are located nearby one another, differences in their measurements of the raw environmental signal will have caused occasional bit errors in the generated bit sequences due to clock jitter, low signal-to-noise ratio, and other sensor variations.
    
    \item \textbf{Key Reconciliation:} 
    Key reconciliation is the process by which two nearby devices exchange messages with one another over a public channel to resolve bit differences in their bit sequences.
    Generally, keys need to be within 20\% bit error rate to authenticate~\cite{kyuin-reconciliation}.
    After key reconciliation stage, both devices will hold identical authentication keys that can be used to symmetrically encrypt data or validate their identities to one another.
\end{enumerate}

\subsection{Threat Model}
\label{sec:threatmodel}


{\system} targets applications in which several headless IoT devices want to autonomously form a network with one another using ZIPA to establish mutual trust.
We assume that legitimate devices are all located inside some physically-secured space such as an office, house, or apartment.
We call this space the legitimate environment.
Individual devices may have limited computation power---IoT devices, mobile phones, or bluetooth accessories all have microcontrollers or low-power CPUs.

Previous ZIPA threat models assume that adversaries have no access to environmental context.
In reality, we demonstrate that although the attacker cannot measure a perfect version of the legitimate environmental context, they can record an approximation of the context by measuring context near a shared wall.
A few common examples of this scenario are:

\vspace{-\topsep}
\begin{itemize}[leftmargin=*]
    \item \textbf{Office Suite Neighbor:} a tenant of an office suite wishes to gain access to the ZIPA network in a neighboring suite.
    \item \textbf{Trespasser:} an attacker covertly places a recording device near a window of a home.
    The recording device measures an approximation of the environmental context that propagates through the building's walls and windows.
    \item \textbf{Apartment Party:} An attacker in a neighboring apartment records audio from high-volume music generated during a party.
    
\end{itemize}
\vspace{-\topsep}

SyncBleed's threat model explicitly assumes that attackers can measure approximate context through permeable walls.
Adversaries under our threat model can intercept all public (unsecured) wireless transmissions made by the legitimate devices inside the legitimate environment.
Although they do not have access to the same ambient environmental signals inside the space, they can measure the environmental context nearby, which is only slightly different from the contexts of legitimate environment as illustrated in \Cref{fig:aliceandbob}.
Adversaries can also masquerade as legitimate devices by initiating authentication. 
Additionally, although the legitimate device's computation power may be limited, we assume the adversary has much more computation power at its disposal (eg. GPUs).

The adversary's goal is to gain access to the network of legitimate devices either by (a) passively listening to wireless communications and environmental signal to learn their shared key or by (b) actively masquerading and participating in the protocol as a legitimate device.
To learn the shared key, the adversary passively snoops the public wireless channel for messages exchanged by legitimate devices.
The information it learns by listening to the channel may be used to form an estimate of the key.
In the masquerading strategy, the adversary impersonates a legitimate device, acquiring measurements of ambient environmental signals and following the protocol to establish a key.
Using a combination of imperfect measurements of the environmental signal and messages exchanged with legitimate devices, the adversary negotiates a key and joins the network.

\subsection{SyncBleed}
\label{sec:infoleak}
In this section, we describe a passive attack on existing ZIPA systems using the information leaked during the synchronization stage.
We call this type of attack \textbf{SyncBleed} because information from ZIPA synchronization messages is bled to nearby passive observers.
SyncBleed is undetectable: an adversary only needs to listen to synchronization packets on a public wireless channel and use local computation to narrow the key space and identify the key.
Other than listening, no interaction with other devices on the network is required.

\subsubsection{Description of SyncBleed}

SyncBleed's threat model assumes that adversaries have physical access to rooms adjacent to the legitimate space.
Adversaries in neighboring apartments, offices, or public hallways can easily carry out SyncBleed.

We use audio as our environmental signal and evaluating the popular Schurmann and Sigg~\cite{Schurmann-TMC13} bit quantization algorithm, but other signal types also work as well, as we demonstrate in \S \ref{sec:eval}.

A legitimate device equipped with a microphone is placed inside an office (depicted in \Cref{fig:prototype} (c)).
An adversarial device, equipped with the same microphone, is placed outside the office door. 
Inside the office, a YouTube video~\footnote{\url{https://youtu.be/xv0YoL57mUA}} is played to mimic a environment with a conversation.
Outside the office, the adversarial device hears a distorted version of that audio but does not contain enough common-mode information to authenticate.

The adversary, situated outside the legitimate space, records a distorted version of the legitimate environmental signal.
The physical barriers between the legitimate space and the adversarial space do not completely mute the signal: instead they act as a kind of filter that selectively muffles some frequency components.
The adversary also snoops synchronization messages---broadcast over a public channel---as its source of audio from inside the legitimate space.
The adversary estimates the filter's transfer function by comparing signal from inside the legitimate space to signal measured in the adversarial space.
The adversary then applies the inverse transfer function to its audio recording to estimate the signal in the legitimate space.


Without noise, we could compute the transfer function $H(\omega) = L(\omega)/A(\omega)$, where $L(\omega)$ is the FFT of the legitimate audio from the synchronization message and $A(\omega)$ is the FFT of the adversary's audio.
But direct computation of $H(\omega)$ tends to amplify noise, especially at higher frequencies.
Instead, we train a simple generative DNN with four hidden layers on the FFT of the adversarial microphone's audio and with the target being the FFT of the legitimate microphone's audio.
The DNN is trained on 1024 synchronization messages using a mean squared error loss function.
We use this ML model to approximate the transfer function between the legitimate and adversary, and attempt to generate bit sequences from the audio signal corrected by the transfer function.

Once we have estimated the transfer function, we can reconstruct the signal inside the legitimate space from muffled measurements taken outside: $E(\omega) = \frac{M(\omega)}{H(\omega)}$.
$E(\omega)$ is the FFT of the adversary's estimate of audio inside the legitimate space, $M(\omega)$ is the FFT of the muffled signal measured just outside the legitimate space, and $H(\omega)$ is the estimated transfer function.
We then inverse transform $E(\omega)$ to reconstruct the estimated time-domain samples inside the legitimate space and apply the Schurmann \& Sigg bit generation algorithm, comparing estimated bits of the attacker to bits generated by a device inside the legitimate space.

Without knowledge of bits generated by a legitimate device, we can passively determine if the adversary's key is correct by applying Reed-Solomon key reconciliation (detailed discussion in \S\ref{sec:algo}).
Key reconciliation can be computed locally without exchanging any messages with legitimate devices.
If key reconciliation succeeds, we are likely to have found the correct key.
If it fails, we can use a brute force attack to find a key that is close to the estimated key in Hamming distance that can be decoded with Reed-Solomon.



\subsubsection{Acquiring Training Data}

SyncBleed needs training data to estimate the transfer function between the attacker and the legitimate space.
We acquired this training data by placing an attacking device outside the legitimate space in our testbed and recording audio and synchronization messages generated by legitimate devices.
The legitimate devices in our experiment were continuously rotating their keys once per minute, and it took about 17 hours to collect 1024 samples of training data.

It is also possible for the attacker to induce a ZIPA network to produce training data by transmitting a pairing request to the legitimate devices.
We also tested this active technique of generating training data.
Using  the active method, we acquired 1024 samples of training data in less than five minutes.
The drawback of actively generating training data is that high-frequency pairing requests may alert the legitimate devices to the attack.

We think these are both realistic vectors for attack as historically adversaries have been willing to be within close proximity to their targets for long periods of time to compromise them.
Tools like \texttt{aircrack-ng}~\cite{aircrack-ng} have been used to learn the key for WEP-encrypted WiFi networks from logs of encrypted packets~\cite{frag-attack}.
This takes advantage of the attacker's proximity to the network and their willingness to collect data for a longer period of time.
This exploit is a genuine security vulnerability, and it is one of the main reasons that nearly all WiFi networks now use WPA encryption.
Our threat model for \textsc{Trevor} and SyncBleed similarly assumes that adversaries may have physical access to snoop synchronization messages and record a distorted version of environmental context within close proximity.

\subsubsection{Results of Attack}

\textbf{Training \& Attack on Same Physical Space}
In the default attack mode, an adversary places a listening device nearby the legitimate space.
First, the adversary gathers training data from the target ZIPA system by eavesdropping synchronization packets.
Then the adversary uses the data to train a machine learning model to approximate the transfer function that separates it from the legitimate network.

\begin{figure}
    \centering
    \includegraphics[width=\hsize]{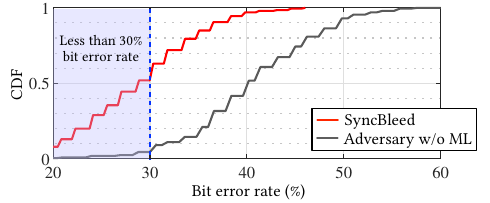}
    \caption{Synchronization messages give an adversary enough information to generate bit sequences that with low bit error rate. Plot shows a CDF of bit error rate of the bit sequences generated by the adversary. Adversary can generate bit sequences with bit error rate below 30\%, which is low enough to masquerade as a legitimate device after reconciliation.}
    \label{fig:attack}
\end{figure}

\Cref{fig:attack} illustrates the CDF of bit error rates between 200 bit sequences generated by the adversary without using SyncBleed against that of an adversary using SyncBleed.
The adversary without using SyncBleed exhibits a mean bit error rate of 41.7\%.
With SyncBleed, the adversary was able to achieve a mean bit error rate of 28.7\%.
Using SyncBleed, an adversary can generate keys with less than 30\% bit error rate more than 50\% of the time.
This attack is simple enough to be carried out by anyone with basic knowledge of neural networks and signal processing.
Worse yet, the adversary's estimate of the transfer function does not expire with the key---it can be used in the future to generate new keys, making key rotation useless for the legitimate devices.

\textbf{Training Data Reused on Attack on a Different Room in Same Building}
In an alternative attack mode, the adversary trains an ML model in one physical space, then re-uses the pretrained model to attack a ZIPA network in a different space.
The training and attack are carried out on two different rooms in our department's office building.
The rooms have different shapes, but they are built using the same construction materials and techniques.
Both rooms have doors made from solid wood with drywall and fiberglass insulation (for soundproofing) separating the rooms.
The adversary re-using the pretrained model achieves a mean bit error rate of 50.1\%---effectively the same as an unassisted adversary.
This illustrates the pretrained model is not capable of generalizing to other rooms.
This result makes sense because SyncBleed approximates the transfer function for a particular room that is based on the room geometry, building materials, and the positions of the legitimate and adversary devices.

Despite being unable to reuse training data across rooms, we think the scenario where the adversary trains and attacks on the same physical space is reasonable because IoT devices are generally stationary.

\subsection{Shift Intolerance of Existing ZIPA Protocols}
\label{sec:shiftintolerance}


\begin{figure}
    \centering
    \includegraphics[width=\hsize]{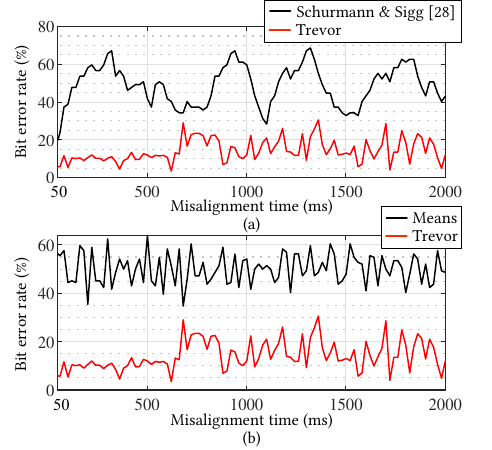}
    \caption{The Means and Schurmann \& Sigg bit quantization algorithms produce high bit error rate between bit sequences that are misaligned when applied to conversation audio. }
    \label{fig:shiftintolerance}
\end{figure}

The two popular ZIPA bit quantization algorithms are (1) Schurmann and Sigg ~\cite{Schurmann-TMC13} and (2) ``means.''
The ``means'' algorithm directly extracts bits from the time-domain samples of an environmental signal, while Schurmann and Sigg's algorithm extracts bits from the frequency spectrum of the signal.
Both suffer from bit errors when the original signal is shifted in time.

In \Cref{fig:shiftintolerance}, we illustrate how small time misalignments can lead to high bit error rates in two popular ZIPA bit quantization algorithms using conversational audio. 
Using synchronously recorded audio from two nearby legitimate devices, we shifted the two buffers in software and generated keys using Means, Schurmann \& Sigg, and {\system}.
The x-axis is the amount of misalignment in time, and the y-axis shows the bit error rate between two legitimate devies.
Even a small misalignment in the signal results in extremely high bit error rates in bit sequences produced by existing bit generation algorithms.

\begin{figure}
    \centering
    \includegraphics[width=\hsize]{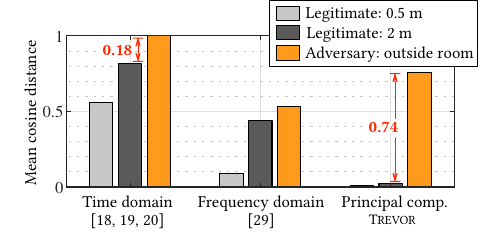}
    \caption{Mean cosine distance between 1 second of
  conversation collected on devices with 100~ms of misalignment times.}
    \label{fig:noisedistance}
\end{figure}

To understand why this happens, we evaluated the similarity of the underlying signals being used for the two popular ZIPA bit quantization algorithm.
The data we analyzed is audio of a conversation recorded in several locations inside an office in our department (depicted in \Cref{fig:prototype}).
We recorded the same audio in four locations: one reference, a second 0.5~m from the reference, a third 2~m from the reference, and a fourth outside the room with the door closed.
We took all recordings simultaneously.

Except for the reference, we iteratively shifted each audio stream in time by 5,000 samples ($\sim$100 ms) and computed cosine distance between the shifted audio and the reference recording at each shift.

We chose a shift time of $\sim$100 ms because it is a bit longer than the worst-case round trip time of the WiFi network we are using to communicate.
Since our goal is to completely avoid time synchronization among ZIPA devices, we need our protocol to be flexible to misalignments that are at least as big as one RTT.

We then computed the mean of the cosine distance over the total shift amount.
The leftmost cluster of bars in \Cref{fig:noisedistance} shows the mean cosine distance between the shifted time-domain audio and the unshifted reference recording for each microphone location.
In the plot, the large mean cosine distance indicates that the shifted audio is dissimilar from the reference.

The middle cluster of bars in \Cref{fig:noisedistance} shows the mean cosine distance between the FFT of the shifted audio and the FFT of the unshifted reference.
There is more similarity between the FFT of the shifted audio and the time domain, but the distance is still not as low as it could be.
Of particular concern is that the legitimate 2 m recording looks similar in cosine distance to the adversary even though it is in the same room as the reference.
This is because the frequency spectrum encodes a lot of noise from the room, especially for recordings that were taken far away from each other.

To separate recordings that were taken in the same room from those that were taken outside the room, we need to choose only the frequency components which carry the common-mode signal.
In the rightmost cluster of bars in \Cref{fig:noisedistance}, we show the mean cosine distance between the four dominant principal components of the FFTs of each recording.
Principal component analysis does the hard work of choosing the important frequency domain components that that carry the most information about the audio stream.
The plot shows that the principal components of the two legitimate devices are very close to the reference in cosine distance, while the adversary is far away.

\section{The {\system} Algorithm}
\label{sec:algo}

\begin{figure*}[t]
    \centering
    \includegraphics[width=\hsize]{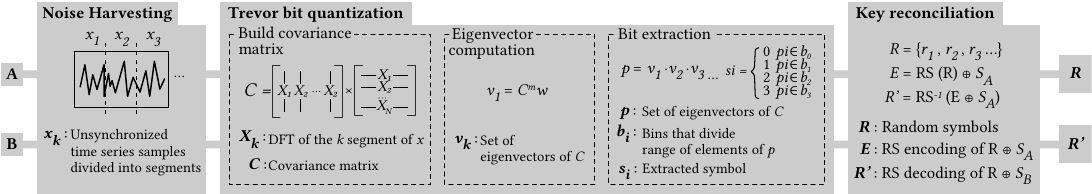}
    \caption{Pipeline used by {\system} for establishing a key.}
    \label{fig:protocol}
\end{figure*}

In this section, we discuss our design of an algorithm and protocol for establishing a shared key between two untrusting devices \textbf{A} and \textbf{B} that are within radio transmit range of one another.
Although devices \textbf{A} and \textbf{B} are within radio transmit range of one another, adversaries are able to listen and transmit messages.
\textbf{A} and \textbf{B} use {\system} to prove to one another that they are located in the same physical space and therefore legitimate.
This explanation assumes that devices \textbf{A} and \textbf{B} are using ambient audio as the basis for building a key, but {\system} also works other kinds of signals.
\Cref{fig:protocol} illustrates the process.

\textbf{Warmup and Sampling}
To begin the process, one device transmits an initiation message to the other.
After initiation, both devices begin sampling ambient audio from their microphones, collecting data for several seconds.
Uncertainty in the wireless channel's latency means that there will be some offset between times when the two devices begin sampling.

\textbf{Build Observation Matrix}
The sequence of time-domain audio samples is then subdivided into blocks of length $d$.
Each block of time-domain samples is then Fourier transformed, discarding phase.
Frequency components of the Fourier spectrum are binned together to build a coarse-grained estimate of the frequency spectrum.
Binning reduces the effects of measurement error and time domain shifts between nearby legitimate devices.
The remaining magnitude information is shift-invariant for shifts that are small compared to the buffer length.
Adversaries located outside the physically-secured space will measure substantially different frequency spectra because walls and doors will selectively muffle frequencies.
The Fourier transformed blocks are arranged as rows of a data matrix $\mathbf{X}$:

\begin{center}
\begin{math}
      \mathbf{X}=
      \left[
      \begin{array}{ccc}
      \horzbar & X_1 & \horzbar \\
      \horzbar & X_2 & \horzbar \\
                    &    \vdots     &\\
        \horzbar & X_{\frac{n}{d}} & \horzbar\\
      \end{array}
      \right]
\end{math}
\end{center}

{\system} discards the first frequency bin, which usually contains some high-amplitude low-frequency components that are common to many kinds of signals.
Low-frequency high-amplitude frequency components tend to dominate the eigenvectors without providing unique context.

The number of rows in $\mathbf{X}$ is controlled by the amount of audio acquired during warmup and sampling.
In general, more rows in $\mathbf{X}$ give better pairing reliability for legitimate devices.

   

\textbf{Generate Covariance Matrix}
The Fourier transformed observation matrix $\mathbf{X}$ is used to calculate the covariance matrix $\mathbf{C} = \mathbf{X}^T\mathbf{X}$ between each frequency component.
If the original time-domain audio has a stable magnitude spectrum over time, $\mathbf{C}$ will capture the overall statistical structure of the signal. 
Shifts in the original time-domain signal will not propagate to significant differences in $\mathbf{C}$ because the rows of $X$ do not vary much under small time shifts.


\textbf{Extract Dominant Eigenvectors}
From $\mathbf{C}$, we extract 4-8 dominant eigenvectors by repeatedly using the power method followed by deflating the matrix.
Each eigenvector specifies the coefficients of a linear combination of frequency components that maximize the spread in the underlying data.

Although in general eigenvector problems tend to be computationally intensive, in practice it is manageable on a modest CPU.
{\system} computes a covariance matrix on the order of dimension $32 \times 32$ which can comfortably fit in memory even on a microcontroller.

Our power method computations are implemented in C to accelerate eigenvector decomposition on {\system}'s ARM Cortex-A72-based hardware platform (\S\ref{sec:arch}).
We have also implemented a port to run on ARM Cortex-M microcontrollers (\S\ref{sec:cortexm4}).
The eigenvectors computed in this step will be used to extract bits to encode a key, but first they must be corrected for noise from data acquisition.

\textbf{Eigenvector Correction}
Covariance matrices built by devices \textbf{A} and \textbf{B} are slightly different because they are based on audio sampled at different locations in a room.
The resulting eigenstructure of $\mathbf{C_A}$ and $\mathbf{C_B}$ will also be slightly different.
But as long as the underlying audio signals are similar, the dominant eigenvectors will also be similar, while less dominant eigenvectors will have larger errors.
The first 4-8 eigenvectors are usually similar enough to be useful in key generation.
By including more eigenvectors, we can extract more bits at the expense of a higher bit error rate.

Even very similar covariance matrices are likely to have slight differences in their eigenstructure.
The most common problem that arises is that two similar covariance matrices have corresponding eigenvectors that point along the same dimension in opposite directions.
If we can detect this problem, it is easy to correct simply by inverting the sign of every component of the eigenvector.
But without sharing the complete set of eigenvectors between the authenticating devices, this is a very difficult problem to detect.
Since we are working in a high-dimensional space, there is not one preferred direction that we can point all eigenvectors toward.
The problem of eigenvector inversion is more prevalent on lower-order eigenvectors, but it occasionally arises on the first or second dominant eigenvector.

Our solution to this problem is to force the sign of the largest component in absolute value of each eigenvector to be positive.
This is reliable on most covariance matrices with small amounts of error.

A second problem is that the dominance order of some eigenvectors can be swapped.
For example, one eigenvector may be associated with the largest eigenvalue in covariance matrix $\mathbf{C_A}$, while the same eigenvector may be associated with the second largest eigenvalue in covariance matrix $\mathbf{C_B}$.
Inconsistencies like this are well known for classical PCA if the vectors have dimensions larger than the amount of vectors used to build the covariance matrix ~\cite{johnstone2009consistency}. 
Our solution to this is to require the amount of vectors used to build the covariance matrix to be much greater than their dimensions.



\textbf{Generate Bits}
To build a bit sequence from the dominant eigenvectors of the covariance matrix, we append the components of the eigenvectors together into one large array $p$.
We bin the elements of the array into four equally-sized bins that span the range of $p$.
We assign a two-bit symbol $s_i$ to each component of $p$ based on which bin it lives in.

\textbf{Key Reconciliation}
We use the sequence of symbols produced by {\system} to encrypt a randomly-generated key $\mathbf{R}$, and transmit that key from one device to another.

Key $\mathbf{R}$ is encoded with a Reed-Solomon\cite{RSCrypto} error correcting code.
Then the coded symbols produced by {\system} are added to the encoded $\mathbf{R}$ to form an encrypted message that can only be decrypted with a similar sequence of symbols.
$$
E = RS(R) \oplus S_A
$$

The resulting sequence $E$ is then transmitted over a public channel.
The other device then reconstructs an estimate of $\mathbf{R}$ by adding its locally-generated sequence of symbols:
$$
R' = RS^{-1}(E \oplus S_B)
$$

If $S_A \approx S_B$, the Reed-Solomon decoder will correct the bit stream to get $R' = R$, which can be used as an encryption or authentication token.



\section{{\system} Prototype Architecture}
\label{sec:arch}

\begin{figure*}[t]
    \centering
    \includegraphics[width=\hsize]{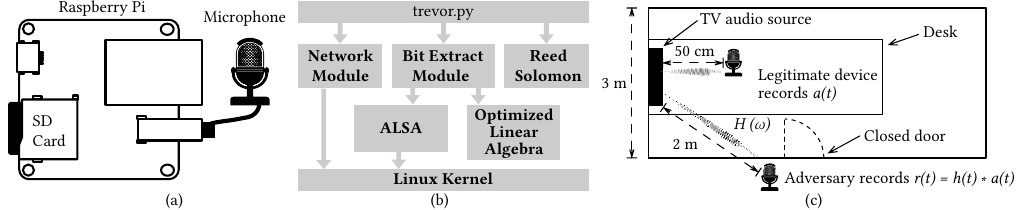}
    \caption{Architecture of our prototype system and testbed: (a) diagram of our Raspberry Pi-based hardware prototype, (b) software architecture of our implementation of {\system}, and (c) a floorplan of the testbed where we evaluated our system.}
    \label{fig:prototype}
\end{figure*}



We built a prototype system to evaluate {\system} on ambient audio in real time.
We demonstrate later that {\system} is a generalizable technique that can be applied to various kinds of environmental signal (other than audio), but we use audio in our prototype system because it is relatively easy to work with, and sensors (microphones) are readily available.

The prototype we describe in this section is intended to mimic the functionality of a mobile phone or IoT device that would use ZIPA scheme to form a network.
Our prototype implementation of {\system} includes the software that generates keys using the algorithm described in \S \ref{sec:algo} and the hardware platform we built to test it.

Nodes in our system are constructed with Raspberry Pi 4~\cite{raspi4} modules running Debian.
Our prototype's hardware capabilities are intended to mimic a mid-range IoT device such as a voice assistant, smart thermostat, or robotic vacuum.

Our hardware prototype is depicted in \Cref{fig:prototype}.
The Raspberry Pi platform is a minimal board running Linux based around an ARM Cortex-A72 CPU running at 1.8 GHz with 2 GB of RAM.
The Raspberry Pi modules connect to each other via a WiFi network that allows them to exchange initiation and key reconciliation messages.
Each Raspberry Pi has a HyperX SoloCast USB microphone to record audio~\cite{solocast} sampling at 48kHz.
The configuration of nodes in our evaluation testbed is depicted in \Cref{fig:prototype}(c).

{\system}'s software architecture is shown in \Cref{fig:prototype}(b).
the main script that orchestrates the pairing process (\S \ref{sec:algo}) is implemented in \texttt{trevor.py}.
Our prototype sends an initiation message to begin pairing and a key reconciliation message (labelled $E$ in \Cref{fig:protocol}) neither message leaks information about the audio (see \S \ref{NIST_RES} for details).
{\system}'s bit extraction module implemented in \texttt{be.py} reads audio data from the microphone, builds an FFT data matrix, computes the covariance matrix, extracts eigenvectors, and generates a key.
Our prototype implements a C-language linear algebra library optimized to run on the Raspberry Pi platform to compute matrix multiplications and eigenvector decomposition.


\section{Evaluation}
\label{sec:eval}


In this section we investigate how {\system} performs under various input conditions.
We used our prototype system to acquire audio data inside our department's office building.
Audio in all of our experiments was generated by playing YouTube videos of conversation~\cite{conversationaudio} and music~\cite{musicaudio} in the lead author's office.
We placed several of our devices throughout the office space (depicted in \Cref{fig:prototype}) and recorded the audio coming from a speaker in the room.
We also tested with audio from the public DEMAND~\cite{demand} dataset, which is a recording of ambient audio in a home.

Some of the experiments in this section require perfectly aligned audio buffers to study how shifts affect properties of the keys.
For those experiments, we used a single laptop to log synchronized audio data from multiple microphones simultaneously.
Other experiments study how well {\system} performs in a realistic deployment scenario, with multiple devices attempting to pair with each other.
For those experiments, we used the platform shown in \Cref{fig:prototype}, which has a network connection and a microphone.
Audio collected by multiple nearby devices cannot be perfectly synchronized.

Two important figures of merit that we study in detail are \emph{bit error rate} and \emph{pairing success rate}.
The bit error rate of a pair of bit sequences generated by two pairing devices is the proportion of bits that do not match.
Bit sequences are then used to generate authentication keys via key reconciliation.
The pairing success rate is the proportion of authentication attempts that succeed---when the bit sequences are similar enough to produce a shared key.
We report both bit error rate and pairing success rate as percentages.
Legitimate devices should produce bit sequences with low bit error rate and high pairing success rate.
An adversary pairing with a legitimate device should have bit error rate of $\sim 50\%$ and a pairing success rate of zero.
Table \ref{tab:ber} compares {\system}'s bit error rate to other ZIPA systems.

\subsection{Resistance to Shifting}
\label{sec:shifteval}

\begin{figure}
    \centering
    \includegraphics[width=\hsize]{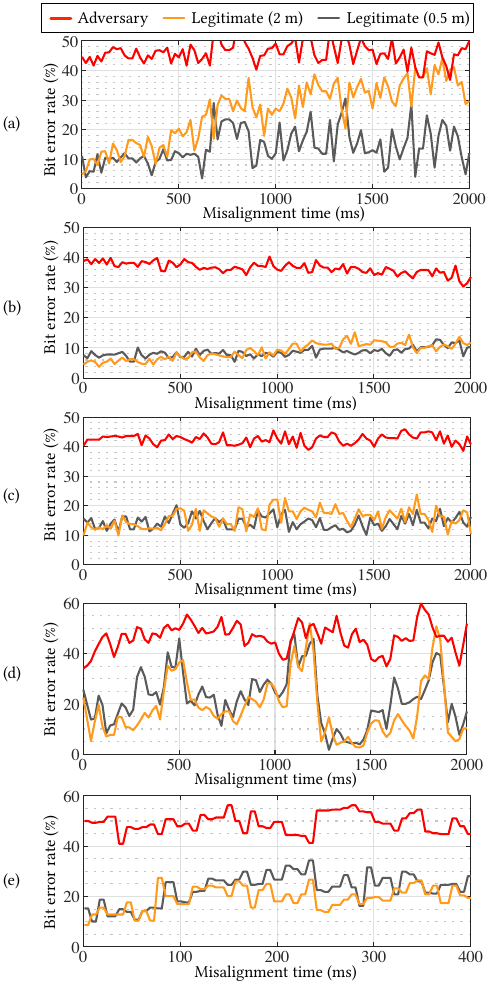}
    \caption{Bit error rate as a function of shift amount for (a) conversation audio, (b) music audio, (c) DEMAND living room ambient audio, (d) voltage, and (e) electromagnetic.}
    \label{fig:bit_error_rate}
\end{figure}

\begin{table}
    \caption{Bit error rate for other ZIPA systems. Our prototype of {\system} achieves comparable bit error rates without exchanging synchronization messages.}
    \label{tab:ber}
    \begin{tabular}{l r}
    \toprule
    \textsc{System} & Bit Error Rate\\
    \midrule
    \textsc{VoltKey}~\cite{voltkey} & 5\% \\ 
    \textsc{AeroKey}~\cite{aerokey} & 9\% \\ 
    \textsc{Secret From Muscle}~\cite{Yang-EMG} & 8\% \\ 
    \textsc{ivPair}~\cite{ivpair} & 5\% \\ 
    \textsc{ProxiMate}~\cite{proximate} & 3-50\% \\ 
     \textbf{\textsc{Trevor}} & \textbf{5-10\%} \\ 
     \bottomrule
    \end{tabular}
\end{table}

Here we study how well the keys that {\system}'s algorithm generates agree with one another when the audio buffers are not well aligned.
This experiment simulates a scenario in which three devices mutually establish one common key derived from shared audio.

The goal of {\system}'s shift resistance is to permit sample buffers measured by authenticating devices to be shifted relative to one another by at least one round trip time on the wireless network.
This permits pairing devices to start the bit generation process without carefully synchronizing with one another.
In our experience working with other similar systems, network delays of $\sim$50 ms are typical for ZIPA devices.
The bit agreement rates for the legitimate devices increased after about 2 seconds of misalignment, which is much longer than the typical round trip time of wireless networks.
\textbf{{\system} performs well for misalignment times that are typical of ZIPA systems.}


The \emph{reference device} is assumed to be a legitimate device inside the physically-secured room that the other three devices are trying to authenticate with.
The \emph{legitimate devices} are nearby the reference device in the physically secured space.
The \emph{legitimate-medium devices} are further from the reference device at approximately two meters away but also in the secure space.
Outside of the physically secured space behind a closed door, the \emph{adversary} is recording audio and attempting to pair with the reference device.
In our configuration, we study three pairs of devices: reference-legitimate, reference-legitimate-medium, and reference-adversary.

The results shown here use synchronous audio data gathered by a laptop with multiple microphones.
Our experiments are done over audio, voltage, and electromagnetic datasets.

\textbf{Audio}
Audio data used in this section comes from synchronous recordings taken in the lead author's office as well as from the DEMAND\cite{demand} dataset.
The DEMAND audio dataset is a collection of 16 microphones gathering ambient audio from the living room from a person's home for five minutes.
The adversary is using ambient audio gathered from the kitchen nearby.
These events do not occur at the same time.
Audio acquired in our offices was recorded from YouTube videos.
Audio was acquired at a sampling frequency of 48 kHz, so the maximum shift on the x-axis represents two seconds of time delay.

\textbf{Voltage}
Voltage data was gathered from our department's offices using a four-channel oscilloscope.
Each channel of the oscilloscope was connected to the 120VAC mains power through a notch filter that attenuated the 60 Hz fundamental frequency.
We attenuated the 60 Hz component because it has a high amplitude and it is common to legitimate and adversary, thus not a good test of whether a devices is within the physically controlled space.
The reference and legitimate devices were plugged in to two nearby plugs that share a common breaker on one floor of our office building, and the adversary was plugged in to a plug on a different floor (served by a different breaker).
Voltage data was acquired at a sampling frequency of 31.25 kHz, so the maximum shift on the x-axis represents 20 seconds of time delay.

\textbf{Electromagnetic}
Electromagnetic data was gathered from our department's offices using a four-channel oscilloscope with each channel connected to a 10 cm antenna.
The reference and legitimate antennas were placed 1cm apart.
The medium-distance antenna was placed 20cm away.
And the adversary antenna was placed 1.5m from the reference.

For our collected audio datasets \Cref{fig:bit_error_rate}(a) and \Cref{fig:bit_error_rate}(b) we can observe several underlying patterns.
The bit error rate of the music data is consistently under 15\% across the entire two seconds of sample shifts.
Whereas the conversation data initially is under 20\% bit error but eventually rises to 40\% as the samples shift.
We believe that this is caused from music having more unique repetitive patterns in the frequencies that stand out in the statistics of the signal.
The main takeaway from these two plots is that when more shared events are present, even when samples are shifted, the devices have a much lower bit error rate.

DEMAND lacks events which makes it an interesting ambient audio dataset with room specific activities. 
The bit error rates are depicted in \Cref{fig:bit_error_rate}(c).
As one can see, the bit error rates are consistently below 20\%.
The main takeaway \Cref{fig:bit_error_rate}(c) is that ambient audio retains low bit error rates, even in the absence of distinct events.

In \Cref{fig:bit_error_rate}(d), large shift amounts cause instability in the eigenstructure of the covariance matrix.
The some of the eigenvectors in the covariance matrices computed by the two legitimate devices point along nearly the same dimension but in opposite directions.
Our method for correcting the directions of the eigenvectors fails in cases where the sample buffers are shifted by large amounts, causing bit error rates to alternate between 30\% and 70\%.
If we had a more reliable method to correct the eigenvector directions, we could further improve shift resistance.

In \Cref{fig:bit_error_rate}(e), electromagnetic data maintains around below 20\% bit error rate for about 100 ms.
After that, it starts to consistently raise above 20\% bit error but still maintains a low bit error rate between 20\% and 30\%.

\subsection{Pairing Success Rate}

\begin{table}
    \caption{Pairing success rate of {\system}.}
    \label{tab:psr}
    \begin{tabular}{lccc}
    \toprule
                         & \multicolumn{3}{c}{Mean Pairing Success Rate}\\
    \textsc{Signal Type} & Legitimate & Medium & Adversary \\
    \midrule
    Conversation Audio & 100\% & 87\% & 0\% \\ 
    Cooking Audio   & 100\% & 70\% & 0\% \\ 
    Voltage         & 80\% & 20\% & 0\% \\ 
    Electromagnetic & 92\% & 95\% & 0\% \\ 
     \bottomrule
    \end{tabular}
\end{table}

To understand the pairing success rate of {\system}, we applied Reed-Solomon key reconciliation (\S \ref{sec:arch}) with a symbol length of 8 bits.
Because there is a wide separation between the bit error rate of keys produced by legitimate devices and those produced by adversaries (\Cref{fig:bit_error_rate}), Reed-Solomon can almost always distinguish between legitimate and adversary.
We parameterized our Reed-Solomon key reconciliation module to error correct keys with no more than 12.5\% bit error rate.
For most of the audio data in \Cref{fig:bit_error_rate}, we get 100\% pairing success rate.
\Cref{tab:psr} summarizes the pairing success rate of {\system} for the modes of environmental context we studied.

\subsection{Replay Attack}

The replay attack simulates a scenario in which an adversary has temporary access to a physically-secured area.
The adversary records audio in the physically secured space while it has access, then uses the recorded audio later to generate a key.
This can happen for example if the adversary is a contractor who enters a building to conduct a repair.
The adversary may return to the site the next day without physical access to the building and attempt to use audio recordings from within the building to generate a key.

\begin{figure}
    \centering
    \includegraphics[width=\hsize]{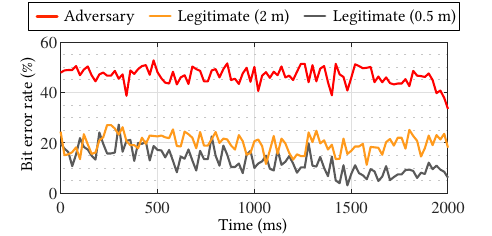}
    \caption{Replay attack bit error rate: legitimate devices use ambient cooking audio. Whereas, an adversary uses old conversation audio recorded from the reference microphone.}
    \label{fig:replay_attack_bit_error_rate}
\end{figure}

We simulated this scenario by recording audio from two different audio files in the same room.
The two recordings we used were taken on two different days using similar microphone configurations.
\Cref{fig:replay_attack_bit_error_rate} compares the bit error rate as a function of shift amount for legitimate devices and an adversary.
The legitimate devices record audio inside the physically secured space at the same time.
The adversary generates a key from audio recorded earlier in the same room.
The adversary is unable to pair with the legitimate devices because its bit error rate is $\sim 50\%$.
The adversary is unable to pair because the frequency spectrum of its pre-recorded audio does not match the frequency characteristics of the audio measured by the legitimate devices even though the adversary's audio was recorded in the same room.

\begin{figure}
    \centering
    \includegraphics[width=\hsize]{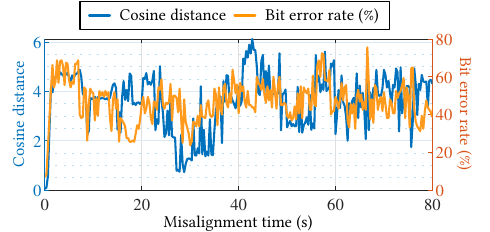}
    \caption{Bit error rate between two legitimate devices as a function of misalignment times.}
    \label{fig:longshift}
\end{figure}

\subsection{Bit Error Rate Generated by our Prototype System}

In this experiment, we generate keys in real time on our prototype implementation.
We set up two pairs of devices in this experiment: legitimate-reference and adversary-reference.
The legitimate and reference devices were both located inside our department's lab with the door closed.
The adversary was located immediately behind the lab's door.
All devices were within 2 meters of one another.

\begin{figure}
    \centering
    \includegraphics[width=\hsize]{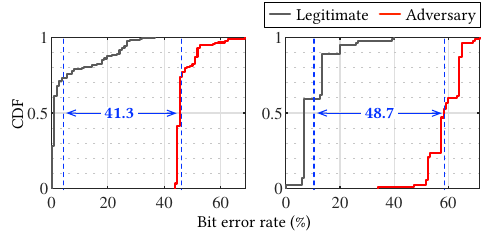}
    \caption{{\system} can distinguish between legitimate and the adversary devices with more than 40\% bit error rate. This figure shows a CDF of bit error rates of bit sequences using (a) conversation audio and (b) cleaning audio.}
    \label{fig:system-conv-cdf}
\end{figure}

We played about 40 minutes of YouTube videos inside of the lab space and generated keys on our adversary in real time.
\Cref{fig:system-conv-cdf} compares CDFs of the legitimate device's bit error rate to the adversary's bit error rate.

{\system} performs well on conversation audio because the signal has a lot of variability.
More specifically, the difference between mean bit error rate of the legitimate and the adversary on conversation audio is 41.3\%. 
Although {\system} has a slightly higher bit error rate on cleaning audio, the bit error rate is well below that of the adversary with the difference of 48.7\%.
It does not perform as well on cleaning audio, because the sound consists of continuous vacuuming, which is much more monotonous than the conversation audio.

\begin{figure}
    \centering
    \includegraphics[width=\hsize]{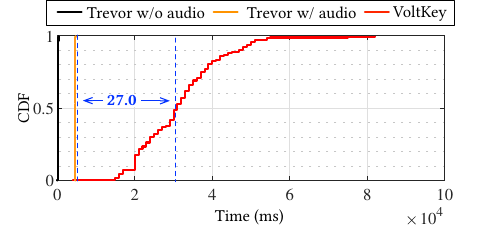}
    \caption{ Evaluation of \system's overall run time on our Raspberry Pi based system.}
    \label{fig:system_clean_cdf}
\end{figure}

\subsection{{\system} Protocol Latency}
\label{sec:time_after_audio}

In this experiment, we evaluate {\system}'s algorithm on the ARM-Cortex A72-based prototype described in \S \ref{sec:arch} as well as on a microcontroller board.
Our goal is to understand how the algorithm performs on realistic application platform.

\subsubsection{Run time on Cortex-A72 Prototype}
Here we evaluate the total end-to-end key generation time of our algorithm running on Raspberry Pi-based prototype, including the time for a pair of authenticators to initiate measure their environmental audio, and compute a key.
{\system}'s protocol uses one short initiation message exchanged between two devices followed by a key reconciliation message to authenticate (\S\ref{sec:algo}).
By comparison to algorithms employed by other ZIPA systems, our protocol exchanges a comparatively small number of messages between authenticating devices.
\textsc{VoltKey}~\cite{voltkey} exchange 11 messages respectively to establish a key, one of which in each protocol is several kilobytes in length.
In practice, exchanging more messages---particularly long ones over unreliable network interfaces---substantially increases the amount of time to establish a key.


{\system} establishes keys far more quickly than \textsc{VoltKey} which show mean run time of around 32~s.
The total time to establish a key is about 5~seconds on average, most of which is taken by recording audio, which is about 27~s faster than the mean run time of \textsc{VoltKey}. 
Not including acquisition of the audio signal from PyAudio, {\system} computes a key in under 10 milliseconds.

\subsubsection{Runtime on Cortex-M4 Microcontroller}
\label{sec:cortexm4}

Here we evaluate the time to compute a key on a Cortex-M4 microcontroller, not including time taken to acquire audio.
We provided a pre-recorded time-domain audio buffer to our algorithm and timed the key generation process on the microcontroller.
Our goal in this experiment is to determine if {\system} is capable of running on a heavily resource-constrained platform.

We implemented the {\system} bit extraction algorithm on a SAME54 Xplained Pro board, which has an ARM Cortex-M4 CPU.
The SAME54 microcontroller~\cite{same54} we used for this experiment has a single-precision floating point unit and runs at 50 MHz.
The platform has 256 kbytes of SRAM and 1 MByte of flash.
We implemented the entire {\system} protocol except for audio sampling and message exchange, which requires an audio input and network interface that are not available on the SAME54 Xplained platform.
Our microcontroller software generates a key from 2.7 seconds of audio with 128k audio samples.
The software we evaluated is a C-language port of the one shown in \Cref{fig:prototype}.
We used ARM's CMSIS-DSP library~\cite{cmsisdsp} for basic linear algebra operations, and we ported our eigenvector decomposition software to work with that library.

\Cref{tab:mcu_eval} shows a summary of {\system}'s performance on the SAME54 microcontroller platform.
Our algorithm takes about 3.2 seconds to compute a key on the Cortex-M4 platform.
Although it takes much longer to compute a key on the microcontroller, it appears that {\system} could be made to work on a full microcontroller-based system.

\begin{table}
    \caption{Evaluation of {\system} running on Cortex-M4 microcontroller without collecting audio.}
    \centering
    \label{tab:mcu_eval}
    \begin{tabular}{l r}
    \toprule
     {\system} Runtime & 3200 ms \\
     Code (flash) size & 161 kbytes \\
     Data (RAM) size & 11 kbytes \\
     \bottomrule
    \end{tabular}
    
\end{table}




\subsection{Randomness of Bit Sequences}\label{NIST_RES}

\begin{table}[t]
\caption{NIST test results of generated bit sequences. (\checkmark indicates pass and \xmark ~indicates failure).}
\centering
\resizebox{\columnwidth}{!}{%
   \centering
   \begin{tabular}{l c c c }
   \toprule
   \textsc{\textbf{NIST Test}} & {\textbf{\system}} & \textbf{\textsc{H2H\cite{h2h}}} & \textbf{\textsc{VoltKey}\cite{voltkey}} \\ 
   \midrule
   \textsc{Frequency} & \checkmark  & \checkmark & \checkmark \\ 
   \textsc{Block Frequency} & \checkmark & \checkmark & \checkmark \\ 
   \textsc{Cumulative Sums} & \checkmark & \checkmark & \checkmark \\ 
   \textsc{Runs} & \checkmark & \xmark & \xmark \\ 
   \textsc{Longest Run} & \checkmark  & \xmark & \xmark \\ 
   \textsc{Rank FFT} & \checkmark & \checkmark & \checkmark \\ 
   \textsc{Non-Overlapping} & \xmark & \checkmark & \checkmark \\ 
   \textsc{Overlapping} & \checkmark & \xmark & \xmark \\ 
    \textsc{Universal} & \xmark & \xmark & \xmark \\ 
   \textsc{Approximate Entropy} & \checkmark & \checkmark & \checkmark \\ 
   \textsc{Serial} & \checkmark & \xmark & \xmark \\ 
   \textsc{Linear Complexity} & \xmark  & \checkmark  & \checkmark \\ 
   \midrule
   \textsc{Pass rate}& 9/12  & 7/12  & 7/12 \\ 
   \end{tabular}
   }
\label{tab:nist}
\end{table}

The generated bit sequences are evaluated for their randomness in Table \ref{tab:nist}.
We evaluated our randomness using the NIST randomness suite which evaluates the safety of random number generation algorithms for cryptographic applications.
For an algorithm to pass more NIST tests would imply the algorithm in question is more secure than one which passes less tests.
{\system} passes 9 of 12 NIST tests, comparable to other ZIPA systems.
The results show that our bits are more random and are unique enough to generate cryptographic keys.

\section{Related Work}

\textbf{Bit Distillation}
Our first area of related work pertains to bit distillation.
Bit distillation algorithms take a bit stream as input and outputs a more random bit stream\cite{west2021moonshine}.
Common uses for bit distillation are applications where the platform is restricted to sampling randomness from a low-entropy source\cite{h2b}.
The problem that bit distillation aims to solve is low randomness found in bit streams.
We differ from bit distillation because we focus on composing time reliant sensor data to matching bit streams.

\textbf{Fuzzy Extractors}
When sampling from an environmental noise source, there are cases where noise gathered from the same scene from two different perspectives does not match.
Fuzzy extractors are algorithms that can account for different perspectives or slightly different readings from the environment to allow for stable use in cryptographic algorithms~\cite{juels1999fuzzy,juels2006fuzzy}.
These algorithms are commonly used for biometric identification~\cite{maltoni2009handbook,sahai2005fuzzy}.
{\system} differs from the work in fuzzy extractors in that it generates a single matching key rather than allowing the key to differ.

\textbf{Zero Involvement Pairing and Authentication (ZIPA)}
Our work is similar to context-based authentication because we rely on shared context to generate a key.
Context-based authentication (CBA) schemes rely on shared entropy seen by at least two devices.
CBA is commonly used in IoT systems\cite{miettinen2018revisiting, fastzip, perceptio}, autonomous vehicles\cite{zeng2019mix}, and smart home security \cite{sreedharan2019securitization} to name a few applications.
Our work differs from common CBA implementations because CBA schemes are reliant on the time domain for synchronous key exchanges~\cite{cba-zipa}. 
Efforts by others are underway to mitigate security compromises that come with ZIPA~\cite{argus, iot-sec}.

\section{Discussion}

\textbf{Privacy:} Direct measurements of the ambient environment divulge private information to eavesdroppers when they are shared on a public channel.
For example, snippets of sensitive audio recordings from inside a locked room could be intercepted by eavesdroppers to learn business secrets, private health information, or personal conversations.
{\system} does not use synchronization messages to establish a key, so it does not need to transmit this private information.

\noindent
\textbf{Security:} Using SyncBleed, synchronization messages can be used to generate valid keys even from outside the physically authenticated space.
In \S \ref{sec:infoleak}, we demonstrated an attack that uses synchronization messages to estimate the channel transfer function between a legitimate device and an attacker outside the physically authenticated space.
We use the channel transfer function to build valid keys from outside the physically authenticated space.
{\system} is not susceptible to this attack because it does not need to exchange synchronization messages to establish a key.

\noindent
\textbf{Speed:} Sharing synchronization messages over a slow network interface slows down the authentication process to the point that it becomes almost unusable.
The problem is especially acute on low-resource IoT devices that would otherwise benefit from ZIPA.
For example, \textsc{VoltKey} platform takes about 30 seconds to establish synchronization.
{\system} authenticates faster than \textsc{VoltKey} and other  ZIPA systems because it does not waste time exchanging sample buffers.



\noindent
\textbf{Trade off Between Shift Invariance and Bit Rate:}
In order to achieve shift invariance, {\system} consumes more environmental data per key bit than existing ZIPA bit generation algorithms.
The result is that our methods can produce fewer key bits per second that existing techniques, which can slow pairing down or generate shorter keys.
But the penalty of slow bit generation is far less than the speedup {\system} achieves by minimizing the synchronization messages that are passed between authenticators.

\section{Conclusion}

In this work, we presented {\system}, an algorithm for extracting bits from correlated environmental signals in ZIPA systems.
{\system} is a mitigation for SyncBleed, an attack on ZIPA systems that infers keys generated inside legitimate spaces from snooped synchronization messages.
To our knowledge, SyncBleed is the first direct attempt to attack ZIPA systems.
{\system}'s algorithm uses principal component analysis to extract correlated common-mode information from an environmental signal and reject noise.
We built a system that implements the {\system} and evaluated our techniques on other data sets to study how it performs when applied to other types of environmental signals.
{\system} achieves bit agreement rates, pairing success rates, and key randomness comparable to existing ZIPA systems while transmitting no synchronization messages which will make ZIPA a practical solution for IoT device authentication.

\section*{Acknowledgements}

This work was supported by the NSF under grant OAC-2107020 and NSA NCAE award H98230-22-1-0306.

\newpage
\bibliographystyle{plain}
\bibliography{refs}

\end{document}